\documentclass[a4paper,11pt]{article}
\usepackage{jheppub}
\usepackage{graphicx}
\usepackage{amssymb}
\usepackage{slashed}
\usepackage{feynmf}
\usepackage{braket}
\usepackage{empheq}
\usepackage{dcolumn}
\usepackage{color}
\usepackage{comment}
\usepackage[font=small]{caption}
\usepackage[font=small]{subcaption}

\title{\boldmath Soft Theorems and Memory Effects at Finite Temperatures}

\author{Divyesh N. Solanki, Srijit Bhattacharjee}
\affiliation{Indian Institute of Information Technology Allahabad (IIITA), Devghat, Jhalwa, Prayagraj-211015, Uttar Pradesh, India.}

\emailAdd{divyeshsolanki98@gmail.com, srijuster@gmail.com}

\abstract{We study the soft theorems for photons and gravitons at finite temperatures using the thermofield dynamics approach. The soft factors lose universality at finite temperatures as the soft amplitudes depend on the nature (or spin) of the particles participating in the scattering processes. However, at low temperatures, a universal behavior is observed in the cross-section of the soft processes. Further, we obtain the thermal contribution to the electromagnetic and gravitational memory effects and show that they are related to the soft factors consistently. The expected zero temperature results are obtained from the soft factors and memories. The thermal effects in soft theorems and memories seem to be sensitive to the spin of the particles involved in scattering. 
}

\begin{document}
\maketitle
\flushbottom

\section{Introduction}
\label{intro}

The infrared (IR) behavior of the scattering amplitudes in gravity and other gauge theories provided many interesting and simple structures. One of the earliest uses of such amplitudes was demonstrated in the bremsstrahlung processes in quantum electrodynamics (QED) where the IR divergences of diagrams like electron vertex correction and tree level diagrams involving a photon in the external legs cancel each other in the limit when the photon's energy tends to zero or in the soft limit \cite{Bloch, Kinosh, LN, Peskin-Schroeder}. Later it was shown, such amplitudes can be cast in terms of only the momenta of the massive scatterers and not involving the soft photons or gravitons. The scattering amplitudes of such soft processes would be characterized by a universal factor known as the {\it soft factor} which is independent of the spin or any other details of the scatterers except their charges and momenta. This fact that the IR amplitudes are described by some universal factors and the fact they are insensitive to the short distance details is known as the (leading) {\it soft theorem}. This feature was further extended by Low and others to include a zero-order term in the energy of soft photons \cite{Low_1954,Low_1958, Goldberger} to the amplitudes. This is known as the subleading soft theorem. Later, Weinberg deduced the Ward identity corresponding to the on-shell gauge invariance for such soft amplitudes involving photons and showed it corresponds to the charge conservation. For the case of scattering involving infrared gravitons, the corresponding Ward identity becomes a statement for global energy-momentum conservation. As a byproduct, the gauge invariance for the soft factor in the case of gravitons requires a universal coupling between gravitons and all kinds of matter fields \cite{Weinberg_1964, Weinberg_1965}, known as the equivalence principle.\\

Recently a remarkable relation between the soft theorems and Asymptotic symmetries in asymptotically flat (AF) spacetimes has been obtained \cite{Strominger}. The asymptotic symmetries are residual symmetries that preserve the boundary structure of an AF spacetime asymptotically. They are equivalent to the large gauge transformations that do not vanish at infinity in gauge theories \cite{Gabai, Strominger_2015, Temple_He, Daniel_Kapec, Strominger_2014, Mitra_2016, Laddha_2015}. For AF spacetimes the analyses by Bondi, Van der Burg, Metzner, and Sachs revealed the emerging symmetries at the null boundary of AF spacetimes are not just Poincare symmetries but a much larger group known as the BMS group \cite{Bondi, Bondi_Burg_Metzner, Sachs}. This BMS group contains the `supertranslations', that constitute an infinite dimensional abelian subgroup of it. Supertranslations act non-trivially on the boundary data and give rise to `memory'. It has been shown that a diagonal subgroup of BMS$^+\times$ BMS$^-$ becomes an exact symmetry of the scattering amplitudes involving soft gravitons in flat space \cite{Strominger_2014}. In fact, the soft photon and soft graviton theorems are just the Ward identities corresponding to the large $U(1)$ gauge symmetries in QED and supertranslation symmetries in gravity  \cite{Daniel_Kapec, Strominger_2015, Strominger}. Recently an extension of the BMS group has been obtained in \cite{Barnich_2010, Barnich_2012}, where a new infinite dimensional symmetry called `superrotation' has been introduced. Superrotations lead to a subleading soft graviton theorem \cite{Cachazo-Strominger} and the angular momentum conservation through the Ward identity. Further, it has been shown, the introduction of superrotation leads to a Ward identity for the gravitational S-matrix corresponding to Virasoro symmetry  \cite{Kapec_2014}. For other developments on soft theorems including celetial holography, see the reviews \cite{Pasterski_EPJC, SAGEX} and references therein.\\

The classical limits of the quantum gravity soft amplitudes are recently studied and it has been shown 
that in that limit the amplitudes are related to the low-frequency component of gravitational waveforms produced by massive scatterers \cite{Sen_2020, Sen_2018}. In fact, the soft theorems have a connection with a purely classical phenomenon known as the memory effect, first proposed by Zeldovich and Polnarev \cite{Zeldovich-Polnarev} and further developed by Braginsky and Thorne \cite{BT} in the context of the scattering of gravitating bodies. The gravitational memory effect can be understood as a DC effect in which a permanent shift takes place between two inertial detectors placed at null infinity after the passage of gravitational waves through them \cite{Strominger_Zhiboedov, Strominger, Tolish_Bieri}. Recently it has been shown, this permanent shift of relative position between the detectors is induced by a supertranslation \cite {Kapec_2014}. In fact,  there is a triangular relationship between Soft Theorem, Asymptotic Symmetry, and Gravitational Memory Effect- all happening in deep infrared region \cite{Strominger}. There is also an electromagnetic analog of the memory effect which can be understood in terms of the ``velocity kick." When electromagnetic waves pass through a moving charged particle, its radiated electric field exerts a kick to the moving charges and produces a memory \cite{Bieri, Pasterski}. The detection of the gravitational memory effect is expected to be achieved at LIGO \cite{LIGO}, LISA \cite{LISA} or in the detector and Pulsar Timing Array \cite{PTA_2010, PTA_2014}.\\

In this note, we study the soft theorems at a finite temperature. Cancellation of IR divergences in thermal QED for soft bremsstrahlung processes had been shown in \cite{Johansson_1986}. The next thing one would be interested to check the soft theorems and the thermal contributions to the soft factors. It is expected that the soft factors in the soft amplitudes involving photons and gravitons would receive thermal corrections. However, at low or zero temperature limits, the expressions of soft factors should reduce to the known results. This has been shown by explicit calculations for soft photon and graviton theorems using thermofield dynamics. These features could have some indications of what we expect for such soft scatterings in a Universe where a $T=2.7K$ ambient temperature exists as a background heat bath. The universal feature of soft factors in the thermal case does not hold because the propagators of fermions and bosons contain Fermi and Bose distributions respectively and they get added to the zero temperature amplitudes with different signs. However, in the limit where the hard particles' energy is much higher than the temperature of the heat bath, both the distributions reduce to Maxwellian and in the cross-section, one would be able to restore the universality. It is also observed the gauge invariance remains intact for the leading-order IR behavior of the soft processes.  We also calculate the contributions of thermal effects on electromagnetic and gravitational memories. Since the soft theorem is not universal at finite temperature, we get two different expressions for the memory effect operator, one corresponds to the scattering of bosons and one corresponds to the scattering of fermions. However, here also we can find out a similar limit where the expressions look the same. Unlike the standard soft theorems at zero temperatures, the soft processes depend on the details of the scatterers at finite temperatures. We also verify the fact that the asymptotic symmetries or 'large' gauge symmetries are exact symmetries of scattering matrices even at finite temperatures. One can establish the soft theorems as a statement of Ward identities by a choice of gauge parameter that contains a temperature dependent part. \\

This paper is organized as follows: In sections (\ref{soft_photon_sec}) and (\ref{soft_graviton_sec}), we study the thermal effects on the leading order soft photon theorem and the soft graviton theorem respectively. In section (\ref{EM_memory_sec}), we determine the electromagnetic memory operator and its action on the asymptotic out-state using the soft photon theorem in the thermofield dynamics formalism. Using the same approach in section (\ref{GM_sec}), we determine the gravitational memory operator and its action on the asymptotic out-state. In section (\ref{WI}) we have discussed the relation between the soft theorems and the Ward identities corresponding to asymptotic symmetries. Finally, we conclude by summarizing our results and discussing some outlooks.

\section{Soft Theorems at Finite Temperatures}
\label{soft_theorem_section}
\subsection{Soft Photon Theorem}
\label{soft_photon_sec}
\noindent To determine the effect of finite temperatures, we study the soft theorems in the thermofield dynamics formalism. In this method, a fictitious system or a tilde system, which is an identical copy of the original system, is introduced to define a thermal vacuum such that the vacuum expectation value of an operator is equal to its ensemble average. The total Hilbert space of the combined system is defined as $\mathcal{H}_T=\mathcal{H} \otimes \Tilde{\mathcal{H}}$, whose bases are $\ket{n}\otimes\ket{\Tilde{n}}$. The operators become doublets which are represented as a $2\times 1$ matrix containing a non-tilde operator and a tilde operator. The non-tilde operator only acts on the non-tilde state $\ket{n}$ and the tilde operator only acts on the tilde state $\ket{\Tilde{n}}$. Thus, the propagator takes $2\times 2$ matrix form consisting of $G_{++}$, $G_{+-}$, $G_{-+}$, $G_{--}$, where $+$ and $-$ corresponds to the non-tilde and tilde vertices respectively.  In our calculation, we consider only a non-tilde operator since it describes the physical part of the whole system. The corresponding element of the matrix-propagator is $G_{++}$, which respectively takes the following form for fermions and bosons \cite{Ashok_Das}-
\begin{equation}
    (\slashed{p}+m) \bigg(\dfrac{i}{p^2-m^2+i\epsilon} -2\pi N_F(p_0) \delta\big(p^2-m^2\big) \bigg),
\end{equation}
\begin{equation}
    \dfrac{i}{p^2-m^2+i\epsilon} + 2\pi N_B(p_0) \delta(p^2-m^2).
\end{equation}
Where $N_F(p_0)=\dfrac{1}{e^{\beta |p_0|}+1}$ and $N_B(p_0)=\dfrac{1}{e^{\beta |p_0|}-1}$ are the respective thermal distribution functions. $p$ is the four-momentum of the particle.\\

\noindent We assume the calculation is done by choosing a thermal bath at rest w.r.t. particles involving scattering. This may lose manifest Lorentz-invariance of the calculation but could have been avoided by replacing $\beta p_{\mu}u^{\mu}$ in place of $\beta p_0$ in the distribution function where $u^{\mu}=(1,0,0,0)$ denotes the four-velocity of the bath \cite{Weldon_1982, Landshoff-Taylor}. \\

\begin{figure}
\centering
\begin{subfigure}{0.45\textwidth}
\centering
\includegraphics[]{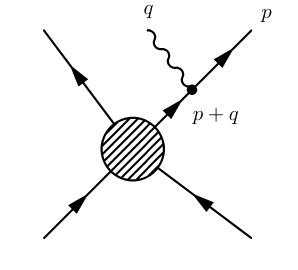}
\caption{}
\label{ffp_out_diagram}
\end{subfigure}
\hfill
\begin{subfigure}{0.45\textwidth}
\centering
\includegraphics[]{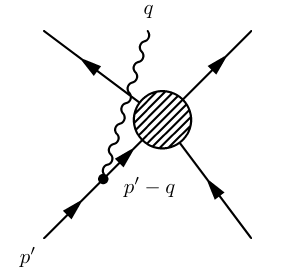}
\caption{}
\label{ffp_in_diagram}
\end{subfigure}
\caption{The above diagrams illustrate the scattering of fermions in which a soft photon is emitted. In fig. (\ref{ffp_out_diagram}), the soft photon is attached to the external leg corresponding to a fermion in the out-state. In fig. (\ref{ffp_in_diagram}), the soft photon is attached to the external leg corresponding to a fermion in the in-state.}
\label{ffp_diagram}
\end{figure}

\noindent First we consider the fermions as the scatterers. In the scattering process, a soft photon is emitted as shown in fig. (\ref{ffp_diagram}). First, we determine the scattering amplitude of the diagram (\ref{ffp_out_diagram}) in which a soft photon of four-momentum $q$ is emitted by one of the fermions of the four-momentum $p$ and the electric charge $Q$.\\

\noindent Using the Feynman rules, the scattering amplitude of the process is written as
\begin{multline}
    M^\mu=\bar{u}(p) (-iQ\gamma^\mu) (\slashed{p}+\slashed{q}+m) \bigg[\dfrac{i}{(p+q)^2-m^2+i\epsilon} -2\pi N_F(p_0+q_0) \delta\big[(p+q)^2-m^2\big] \bigg] \\
\cdot M_{\text{hard}}(p+q).
\end{multline}
Where $M_{\text{hard}}$ is the amplitude of the hard (energetic) part of the process. $M_\text{hard}(p+q)$ also depends on the four-momenta of the scatterers not emitting a soft photon. Here, we have suppressed the explicit dependence of $M_{\text{hard}}$ on these momenta.\\

\noindent In the soft limit, $q\to 0$, we drop the terms $\slashed{q}$ in the numerator and $q^2$ in the denominator and in the Delta function argument. We also write
\begin{align*}
    &N_F(p_0+q_0) \approx N_F(p_0), \\
    &M_{\text{hard}}(p+q) \approx M_{\text{hard}}(p) \equiv M_{\text{hard}}.
\end{align*}
Therefore,
\begin{equation}
    M^\mu = -iQ\bar{u}(p) \gamma^\mu (\slashed{p}+m) \bigg[\dfrac{i}{2p\cdot q +i\epsilon} - 2\pi N_F(p_0) \delta(2p\cdot q) \bigg] M_{\text{hard}},
\end{equation}
which can be written as
\begin{equation}
    M^\mu = Q\bar{u}(p) (2p^\mu - \slashed{p}\gamma^\mu + m \gamma^\mu) \bigg[\dfrac{1}{2p\cdot q +i\epsilon} + 2\pi i N_F(p_0) \dfrac{1}{|2|} \delta(p\cdot q) \bigg].
\end{equation}
Using the identity $\bar{u}(p)(\slashed{p}-m)=0$, we obtain
\begin{equation}
   M^\mu = Q p^\mu \bigg[\dfrac{1}{p\cdot q +i\epsilon} + 2\pi i N_F(p_0) \delta(p\cdot q) \bigg] M_{\text{hard}}.
\end{equation}
Where $\bar{u}(p)$ is absorbed into $M_{\text{hard}}$. Now, we can write down, in a similar way, the amplitude of the second diagram (\ref{ffp_in_diagram}) in which the soft photon of four-momentum $q$ is emitted before the scattering by a fermion of four-momentum $p^\prime$ and the electric charge $Q^\prime$
\begin{equation}
   M^{\prime\mu} = Q^\prime p^{\prime\mu} \bigg[\dfrac{-1}{p^{\prime}\cdot q -i\epsilon} + 2\pi i N_F(p^{\prime}_0) \delta(p^{\prime}\cdot q) \bigg] M_{\text{hard}}. \label{M_prime}
\end{equation}
We may generalize the above expression for $m$ number of fermions involved in the scattering process. Thus, there are $m$ possible ways to attach the soft photon to the external leg in the Feynman diagram. In other words, there are $m$ possibilities of the emission of the soft photon since each fermion can emit it. To determine the total scattering amplitude of the process, we need to sum over all possible Feynman diagrams. Therefore, the total scattering amplitude of the process is written as
\begin{equation}
    M^\mu_{\text{total}} = \sum_{n=1}^{m}\bigg[ \dfrac{Q_n \eta_n p_n^\mu}{p_n\cdot q+i\eta_n \epsilon} + 2\pi i Q_n p_n^\mu N_F(p_{n0})\delta(p_n\cdot q) \bigg] M_{\text{hard}}(p_1, p_2,... , p_m).
    \label{MT_fermion}
\end{equation}
Where $\eta_n = +1$ if the nth charged particle is in the ``out-state", and $\eta_n = -1$ if the nth charged particle is in the ``in-state". $p_1, p_2,... , p_m$ are the four-momenta of the scatterers. The first term in the bracket is the universal soft factor in the quantum field theory, and the second term is the finite temperature contribution to the soft factor.\\

\begin{figure}
\centering
\begin{subfigure}{0.45\textwidth}
\centering
\includegraphics[]{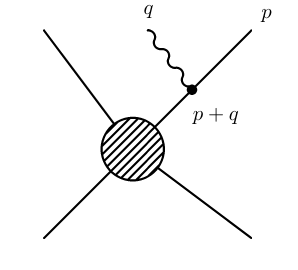}
\caption{}
\label{bbp_out_diagram}
\end{subfigure}
\hfill
\begin{subfigure}{0.45\textwidth}
\centering
\includegraphics[]{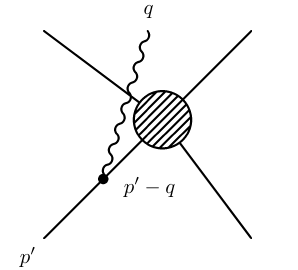}
\caption{}
\label{bbp_in_diagram}
\end{subfigure}
\caption{The above diagrams illustrate the scattering of bosons in which a soft photon is emitted. In fig. (\ref{bbp_out_diagram}), the soft photon is attached to the external leg corresponding to a boson in the out-state. In fig. (\ref{bbp_in_diagram}), the soft photon is attached to the external leg corresponding to a boson in the in-state.}
\label{bbp_diagram}
\end{figure}

\noindent Next, we consider any arbitrary process which now involves the scattering of scalar bosons as shown in fig. (\ref{bbp_diagram}). We now write the scattering amplitude of the diagram (\ref{bbp_out_diagram}) in which a soft photon of four-momentum $q$ is emitted by one of the bosons of four-momentum $p$ and the scalar field charge $Q$.

\begin{equation}
    M^\mu = -i(2Qp^\mu) \bigg[\dfrac{i}{(p+q)^2-m^2+i\epsilon} + 2\pi N_B(p_0+q_0) \delta((p+q)^2-m^2) \bigg] M_{\text{hard}}(p+q)
\end{equation}
which, in the soft limit $q\to 0$, reduces to
\begin{equation}
    M^\mu = Qp^\mu \bigg[\dfrac{1}{p\cdot q+i\epsilon} - 2\pi i N_B(p_0) \delta(p\cdot q) \bigg] M_{\text{hard}}.
\end{equation}
Similarly, one can compute the amplitude of the second diagram (\ref{bbp_in_diagram}) which involves the emission of a soft photon before the scattering. We obtain the same expression as above except for the additional minus sign that appears exactly at the same place as in eq. (\ref{M_prime}). Now, again suppose there are $m$ scalar bosons involved in the scattering. Each scalar boson can emit a soft photon. Thus, adding contributions from all external legs will give us the following expression:
\begin{equation}
    M^\mu_{\text{total}} = \sum_{n=1}^m\bigg[ \dfrac{Q_n \eta_n p_n^\mu}{p_n\cdot q+i\eta_n \epsilon} - 2\pi i Q_n p_n^\mu N_B(p_{n0})\delta(p_n\cdot q) \bigg] M_{\text{hard}}(p_1, p_2, ... , p_m),
    \label{MT_boson}
\end{equation}
where $p_1, p_2, ... , p_m$ are the four-momenta of the scatterers.\\

\noindent It can be seen from eqs. (\ref{MT_fermion}) and (\ref{MT_boson}) that the universality of the soft factor is not maintained at finite temperatures due to the thermal contributions. But, the universality is restored when the hard part dominates over the thermal part of the process. Suppose the energy of the particle associated with the external leg is very high compared to the thermal energy, i.e. $|p_{n0}| \gg \dfrac{1}{\beta}$. 
\begin{equation}
    \therefore N_F(|p_{n0}|) \approx e^{-\beta |p_{n0}|} \approx N_B(|p_{n0}|)
\end{equation}
Thus, we can approximate eqs. (\ref{MT_fermion}) and (\ref{MT_boson}) as
\begin{equation}
    \big(M^\mu_{\text{total}} \big)_F \approx \sum_{n=1}^{m}\bigg[ \dfrac{Q_n \eta_n p_n^\mu}{p_n\cdot q+i\eta_n \epsilon} + 2\pi i Q_n p_n^\mu e^{-\beta |p_{n0}|} \delta(p_n\cdot q) \bigg] M_{\text{hard}}(p_1, p_2, ... , p_m)
    \label{MTF}
\end{equation}
\begin{equation}
    \big(M^\mu_{\text{total}} \big)_B \approx \sum_{n=1}^{m}\bigg[ \dfrac{Q_n \eta_n p_n^\mu}{p_n\cdot q+i\eta_n \epsilon} - 2\pi i Q_n p_n^\mu e^{-\beta |p_{n0}|} \delta(p_n\cdot q) \bigg] M_{\text{hard}}(p_1, p_2, ... , p_m)
    \label{MTB}
\end{equation}
The above expressions indicate the finite temperature contributions to the soft factors are identical except for a sign. Since the scattering cross section is proportional to $|M|^2$ the sign mismatch would go away and a universal behavior would be obtained. \\

\noindent In the following section, we study the thermal effects on the soft factor for gravitons.

\subsection{Soft Graviton Theorem}
\label{soft_graviton_sec}
\begin{figure}
\centering
\begin{subfigure}{0.45\textwidth}
\centering
\includegraphics[]{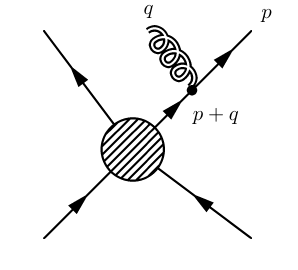}
\caption{}
\label{ffg_out_diagram}
\end{subfigure}
\hfill
\begin{subfigure}{0.45\textwidth}
\centering
\includegraphics[]{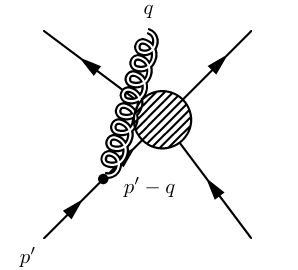}
\caption{}
\label{ffg_in_diagram}
\end{subfigure}
\caption{These diagrams illustrate the emission of a soft graviton in the scattering of fermions. In fig. (\ref{ffg_out_diagram}), the soft graviton is attached to the external leg corresponding to a fermion in the out-state. In fig. (\ref{ffg_in_diagram}), the soft graviton is attached to the external leg corresponding to a fermion in the in-state.}
\end{figure}

\noindent For scattering involving soft gravitons we first consider fermions as the scatterers.  Let, a soft graviton of four-momentum $q$ is emitted by one of the fermions of the four-momentum $p$; which is shown in fig. (\ref{ffg_out_diagram}).\\

The coupling of the graviton with the fermion, i.e. the vertex factor, is given by the following expression:
\begin{equation}
    \dfrac{-i\kappa}{4}\big[(2p^\mu+q^\mu)\gamma^\nu - \eta^{\mu\nu}(2\slashed{p}+\slashed{q}-2m) \big],
\end{equation}
where $\kappa$ is the gravitational coupling constant. Thus, the scattering amplitude is written as
\begin{multline}
    M^{\mu\nu}=\dfrac{-i\kappa}{4}\bar{u}(p)\big[(2p^\mu+q^\mu)\gamma^\nu - \eta^{\mu\nu}(2\slashed{p}+\slashed{q}-2m) \big](\slashed{p}+\slashed{q}+m)\\ \cdot \bigg(\dfrac{i}{(p+q)^2-m^2+i\epsilon} - 2\pi N_F(p_0+q_0) \delta((p+q)^2-m^2) \bigg) M_{\text{hard}}(p+q).
\end{multline}
In the soft limit $q\to 0$, this expression reduces to
\begin{equation}
    M^{\mu\nu} = \dfrac{\kappa}{2} p^\mu p^\nu \bigg(\dfrac{1}{p\cdot q+i\epsilon} + 2\pi i N_F(p_0)\delta(p\cdot q) \bigg) M_{\text{hard}},
\end{equation}
where $\bar{u}(p)$ is absorbed into $M_{\text{hard}}$. The second term in the vertex factor is dropped as the polarization tensors are traceless, i.e. $\varepsilon_{\mu\nu} \eta^{\mu\nu}=0$.\\

\noindent Similarly, if the soft graviton of four-momentum $q$ is emitted by the fermion of four-momentum $p'$ before the scattering, i.e. fig. (\ref{ffg_in_diagram}), the scattering amplitude takes the following form:
\begin{equation}
    M'^{\mu\nu} = \dfrac{\kappa'}{2} p'^\mu p'^\nu \bigg(\dfrac{-1}{p'\cdot q - i\epsilon} + 2\pi i N_F(p'_0)\delta(p'\cdot q) \bigg) M_{\text{hard}}.
\end{equation}
Assuming the scattering of $m$ fermions, we can write the total scattering amplitude as
\begin{equation}
    M^{\mu\nu}_{\text{total}} = \sum_{n=1}^m \dfrac{\kappa_n}{2} p_n^\mu p_n^\nu \bigg(\dfrac{\eta_n}{p_n\cdot q + i\eta_n \epsilon} + 2\pi i N_F(p_{n0}) \delta(p_n \cdot q) \bigg) M_\text{hard}(p_1, p_2,... , p_m),
    \label{M_graviton-fermion}
\end{equation}
where $\eta_n = -1$ if the $n$th fermion is in the in-state, and $\eta_n=+1$ if the $n$th fermion is in the out-state. $p_1, p_2,... , p_m$ are the four-momenta of the scatterers.\\

\begin{figure}
\centering
\begin{subfigure}{0.45\textwidth}
\centering
\includegraphics[]{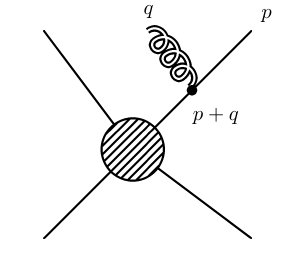}
\caption{}
\label{bbg_out_diagram}
\end{subfigure}
\hfill
\begin{subfigure}{0.45\textwidth}
\centering
\includegraphics[]{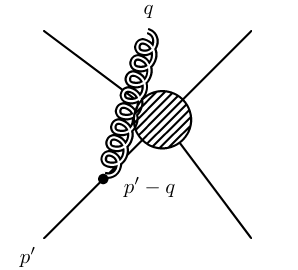}
\caption{}
\label{bbg_in_diagram}
\end{subfigure}
\caption{These diagrams illustarte the emission of a soft graviton in the scattering of bosons. In fig. (\ref{bbg_out_diagram}), the soft graviton is attached to the external leg corresponding to a boson in the out-state. In fig. (\ref{bbg_in_diagram}), the soft graviton is attached to the external leg corresponding to a boson in the in-state.}
\label{bbg_diagram}
\end{figure}

\noindent Now, we consider scalar bosons as the scatterers. A soft graviton of four-momentum $q$ is emitted by one of the bosons as shown in fig. (\ref{bbg_diagram}). The vertex factor for the coupling of the graviton with the scalar boson is given by the following expression:
\begin{equation}
    \dfrac{-i\kappa}{2}\big[p^\mu (p^\nu+q^\nu) + p^\nu(p^\mu+q^\mu) - \eta^{\mu\nu} \big(p\cdot(p+q) - m^2\big) \big].
\end{equation}

\noindent Thus, the amplitude of the diagram (\ref{bbg_out_diagram}) is written as
\begin{multline}
    M^{\mu\nu}=\dfrac{-i\kappa}{2}\big[p^\mu (p^\nu+q^\nu) + p^\nu(p^\mu+q^\mu) - \eta^{\mu\nu} \big(p\cdot(p+q)-m^2\big) \big]\\ \cdot \bigg(\dfrac{i}{(p+q)^2-m^2+i\epsilon} + 2\pi N_B(p_0+q_0) \delta((p+q)^2-m^2) \bigg) M_{\text{hard}}(p+q).
\end{multline}
\noindent Now, we take the soft limit $q\to 0$. And we drop the third term in the vertex factor as $\varepsilon_{\mu\nu} \eta^{\mu\nu}=0$.
\begin{equation}
    M^{\mu\nu} = \dfrac{\kappa}{2} p^\mu p^\nu \bigg(\dfrac{1}{p\cdot q + i\epsilon} -2\pi i N_B(p_0) \delta(p\cdot q) \bigg) M_\text{hard}
\end{equation}
Similarly, if the soft graviton is emitted by the boson of four-momentum $p'$ before the scattering (fig. (\ref{bbg_in_diagram})), we obtain
\begin{equation}
    M'^{\mu\nu} = \dfrac{\kappa'}{2} p'^\mu p'^\nu \bigg(\dfrac{-1}{p'\cdot q - i\epsilon} -2\pi i N_B(p'_0) \delta(p'\cdot q) \bigg) M_\text{hard}
\end{equation}
Now, for the scattering of $m$ bosons, the total scattering amplitude is written as
\begin{equation}
    M^{\mu\nu}_{\text{total}} = \sum_{n=1}^m \dfrac{\kappa_n}{2} p_n^\mu p_n^\nu \bigg(\dfrac{\eta_n}{p_n\cdot q+i\eta_n\epsilon} - 2\pi i N_B(p_{n0})\delta(p_n\cdot q) \bigg) M_\text{hard}(p_1, p_2, ... , p_m).
    \label{M_graviton-boson}
\end{equation}
From eqs. (\ref{M_graviton-fermion}) and (\ref{M_graviton-boson}), we can see, like the soft photon theorem, the universality of the soft factor is not maintained at finite temperatures. However, the four-momentum conservation remains intact as $q_{\mu}. M^{\mu\nu}_{\text{total}}=0$. Consequently, the four-momentum conservation requires the universal gravitational coupling, i.e. $\kappa_1=\kappa_2=...=\kappa_m$, which is the statement of the equivalence principle. The universal behaviour is restored in the cross-section as described for the photon case when the hard part of the scattering process dominates over the thermal part.\\

\noindent In the following sections, we derive the memory operator and its action on the asymptotic out-state by using the soft theorems.

\section{Memory Effect at Finite Temperatures}
\subsection{Electromagnetic Memory Effect}
\label{EM_memory_sec}
\begin{figure}
    \centering
    \includegraphics[scale=0.75]{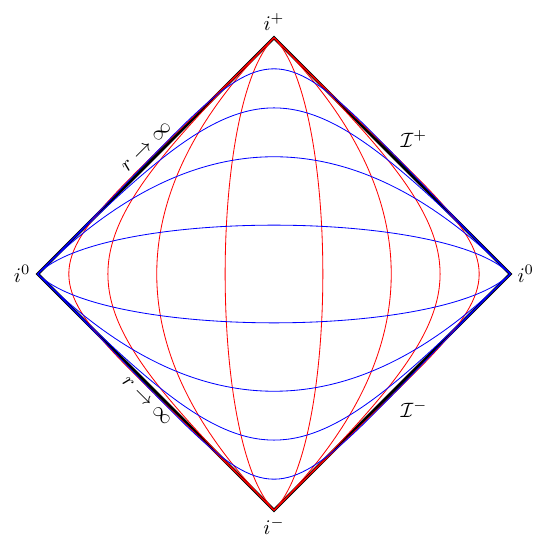}
    \caption{Conformal diagram of the Minkowski space. Blue lines represent constant-t hypersurfaces, red lines represent constant-r hypersurfaces. $i^+$ is future timelike infinity, $i^-$ is past timelike infinity, $i^0$ is spatial infinity, $\mathcal{I}^+$ is future null infinity, and $\mathcal{I}^-$ is past null infinity. All massive particles travel from $i^-$ to $i^+$. Light rays travel from $\mathcal{I}^-$ to $\mathcal{I}^+$.}
    \label{Minkowski_diagram}
\end{figure}

\noindent To study the electromagnetic memory effect one needs to consider the contribution of the electromagnetic gauge field and its derivatives in the radiation using the asymptotic boundary conditions. The memory is an artifact of a non-trivial boundary condition on the gauge field near the future null infinity of the Minkowski space (see fig. (\ref{Minkowski_diagram})).\\

\noindent It will be useful to consider the Minkowski metric in the retarded coordinates $(u,r,z,\bar{z})$:
\begin{equation}
    ds^2=-du^2-2dudr+2r^2\gamma_{z\bar{z}}dzd\bar{z}
\end{equation}
where
\begin{equation}
    \gamma_{z\bar{z}}=\dfrac{2}{(1+z\bar{z})^2}.
\end{equation}

\noindent We have used stereographic coordinates $(z,\bar{z})$ to parameterize the sphere. These stereographic coordinates are related to the spherical coordinates $(\theta,\phi)$ by
\begin{align}
    &z=e^{i\phi}\tan{\dfrac{\theta}{2}},
    &\bar{z}=e^{-i\phi}\tan{\dfrac{\theta}{2}}.
\end{align}
In this parameterization, the unit vector in the cartesian coordinate system can be written as
\begin{equation}
    \hat{x}=\dfrac{1}{1+z\bar{z}}\big(z+\bar{z}, -i(z-\bar{z}), 1-z\bar{z} \big).
\end{equation}

\noindent Now, we discuss about boundary conditions. We work in the retarded radial gauge, $A_r=0$, $A_u|_{\mathcal{I}^+}=0$. Thus, the leading order term in the asymptotic expansion of $A_u$ has the fall-off of $\mathcal{O}(r^{-1})$, and the $A_z$ has the fall-off of $\mathcal{O}(1)$ \cite{Temple_He}. The asymptotic expansions of the gauge fields are
\begin{align}
A_u(u,r,z,\bar{z})=\sum_{n=1}^{\infty}\dfrac{A^{(n)}_u(u,z,\bar{z})}{r^n}, \,\,\,
A_z(u,r,z,\bar{z})=\sum_{n=0}^{\infty}\dfrac{A^{(n)}_z(u,z,\bar{z})}{r^n}.
\end{align}
Using the expression, $F_{\mu\nu}=\partial_\mu A_\nu - \partial_\nu A_\mu$, we obtain the following relations:
\begin{align}
    F_{ur}^{(2)}(u,z,\bar{z}) = A_u^{(1)}(u,z,\bar{z}), \,\,
    F_{uz}^{(0)}(u,z,\bar{z}) = \partial_uA_z^{(0)}(u,z,\bar{z}).
\end{align}
Where $F_{ur}$ corresponds to the radial electric field $\mathcal{O}(r^{-2})$, and $F_{uz}$ corresponds to the radiation field. The radiation fields are responsible for the physical content of the electromagnetic radiation and hence give the electromagnetic memory effect. To understand the electromagnetic memory effect, we consider the following simple example \cite{Bieri}. Suppose a test particle of mass $m$ and electric charge $Q$ is moving with small enough velocity such that we can neglect the magnetic forces acting on it at very large distances. The total electromagnetic radiation reaching at infinity is the integration of the Poynting vector on the asymptotic sphere. The contribution of the radial electric and magnetic field in the Poynting vector has a fall-off of $\mathcal{O}(r^{-4})$. The term $r^2$ from the Jacobian will cancel the fall-off of $\mathcal{O}(r^{-2})$ from the Poynting vector and the resultant expression has the fall-off of $\mathcal{O}(r^{-2})$. And hence at infinity, the radial electric field and magnetic field have no contribution to electromagnetic radiation. The only terms that survive at infinity are the contribution of the radiation fields. Thus, the Lorentz force acting on the test particle is given by
\begin{align}
    &\mathbf{F}=Q \ \mathbf{E}_{\text{rad}} \nonumber\\
       \therefore \ &\Delta\mathbf{v} = \dfrac{Q}{m} \int_{-\infty}^{+\infty} \mathbf{E}_{\text{rad}} \ dt. \label{kick}
\end{align}
The time integration of the radiation field gives the electromagnetic memory effect. Here, we can understand the memory effect on the test particle as the ``velocity kick" given by the above expression (\ref{kick}).\\

\noindent Now, we derive the memory effect operator as introduced in \cite{Pasterski}, and study the finite temperature effects on it. In the free Maxwell theory, the mode expansion of the field operator can be written as \cite{Pasterski, Temple_He}
\begin{equation}
    A_\mu(x)=e\sum_{\alpha=\pm}\int\dfrac{d^3\mathbf{q}}{(2\pi)^3} \dfrac{1}{2\omega} \big[\varepsilon_\mu^{\alpha *}(\mathbf{q}) a_\alpha(\mathbf{q}) e^{iq\cdot x} + \varepsilon_\mu^{\alpha}(\mathbf{q}) a^{\dag}_\alpha(\mathbf{q}) e^{-iq\cdot x} \big],
\end{equation}
where $e^{iq\cdot x}=e^{-i\omega u-i\omega r(1-\hat{q}\cdot\hat{x})}.$ Since we are using thermofield dynamics formalism, we work with the thermal field operator which is defined as
\begin{equation}
    A_\mu(x;\beta)=e\sum_{\alpha=\pm}\int\dfrac{d^3\mathbf{q}}{(2\pi)^3} \dfrac{1}{2\omega} \big[\varepsilon_\mu^{\alpha *}(\mathbf{q}) a_\alpha(\mathbf{q};\beta) e^{iq\cdot x} + \varepsilon_\mu^{\alpha}(\mathbf{q}) a^{\dag}_\alpha(\mathbf{q};\beta) e^{-iq\cdot x} \big].
    \label{A}
\end{equation}
The Lorenz gauge condition requires that the polarization vectors are orthogonal to the four-momentum of the photon. Now, we parameterize the four-momentum of the photon as
\begin{equation}
    q^\mu=\omega(1,\hat{q}),
\end{equation}
where
\begin{equation}
    \hat{q}=\dfrac{1}{1+w\bar{w}}\big(w+\bar{w},-i(w-\bar{w}),1-w\bar{w} \big).
\end{equation}
In this parameterization, the choice of the polarization vectors is
\begin{align}
    &\varepsilon_\mu^{+}(\mathbf{q}) = \dfrac{1}{\sqrt{2}} (-\bar{w},1,-i,-\bar{w} ) \label{varepsilon_mu+} \\
    &\varepsilon_\mu^{-}(\mathbf{q}) = \dfrac{1}{\sqrt{2}} (-w,1,i,-w ) \label{varepsilon_mu-}
\end{align}
such that $\varepsilon_\mu^{\pm}q^{\mu}=0$ and $\varepsilon_\mu^{+} \varepsilon^{-\mu}=1$. From eq. (\ref{A}),
\begin{multline}
    A_\mu(x;\beta)=\dfrac{e}{2(2\pi)^3}\sum_{\alpha=\pm}\int d\omega \ \omega \int d\Omega_{\hat{q}} \big[\varepsilon_\mu^{\alpha *}(\mathbf{q}) a_\alpha(\mathbf{q};\beta) e^{-i\omega u-i\omega r(1-\hat{q}\cdot\hat{x})} \\+ \varepsilon_\mu^{\alpha}(\mathbf{q}) a^{\dag}_\alpha(\mathbf{q};\beta) e^{i\omega u+i\omega r(1-\hat{q}\cdot\hat{x})} \big].
    \label{A1}
\end{multline}
Next, we consider the asymptotic expansion of the above field operator. In the $r\to\infty$ limit, the above integral becomes highly oscillatory. We use stationary phase approximation to approximate the above integral \cite{Strominger}. The phase is stationary at $\hat{q}\cdot\hat{x}=1$, yielding
\begin{equation}
    \int d\Omega_{\hat{q}}f(\mathbf{q})e^{\pm i\omega r(1-\hat{q}\cdot\hat{x})} \xrightarrow[r\to\infty]{} \pm\dfrac{2\pi i}{\omega r}f(\omega \hat{x}).
    \label{stationary_phase_identity}
\end{equation}
Using the above expression (\ref{stationary_phase_identity}) into eq. (\ref{A1}), we obtain
\begin{equation}
    A_\mu(x;\beta)=\dfrac{-ie}{8\pi^2 r}\sum_{\alpha=\pm}\int d\omega \ \big[\varepsilon_\mu^{\alpha *}(\omega\hat{x}) a_\alpha(\omega\hat{x};\beta) e^{-i\omega u} - \varepsilon_\mu^{\alpha}(\omega\hat{x}) a^{\dag}_\alpha(\omega\hat{x};\beta) e^{i\omega u} \big].
    \label{A2}
\end{equation}
We transform the field operator $A_\mu$ into the retarded coordinates $(u,r,z,\bar{z})$. In the new coordinates, we write the $z$-component as
\begin{equation}
    A_z(u,r,z,\bar{z};\beta)=\dfrac{\partial x^\mu}{\partial z}A_\mu(x;\beta),
\end{equation}
where
\begin{align}
    &\varepsilon^{+}_{z} = \dfrac{\partial x^\mu}{\partial z} \varepsilon^{+}_{\mu}(\omega\hat{x}) = 0 \label{varepsilon_z+} \\
    &\varepsilon^{-}_{z} = \dfrac{\partial x^\mu}{\partial z} \varepsilon^{-}_{\mu}(\omega\hat{x}) = \dfrac{\sqrt{2}r}{1+z\bar{z}} \label{varepsilon_z-}.
\end{align}
Therefore,
\begin{equation}
    A_z(u,r,z,\bar{z};\beta)=\dfrac{-i\sqrt{2}e}{8\pi^2 (1+z\bar{z})} \int_{0}^{\infty} d\omega \ \big[a_+(\omega\hat{x};\beta) e^{-i\omega u} - a^{\dag}_-(\omega\hat{x};\beta) e^{i\omega u} \big],
    \label{A3}
\end{equation}
where $a^{\dag}_-$ creates a photon of negative helicity, and $a_+$ annihilates a photon of positive helicity. The RHS of eq. (\ref{A3}) is the leading order term in the asymptotic expansion of $A_z$,
\begin{equation}
    A^{(0)}_z(u,z,\bar{z};\beta)=\dfrac{-i\sqrt{2}e}{8\pi^2 (1+z\bar{z})} \int_{0}^{\infty} d\omega \ \big[a_+(\omega\hat{x};\beta) e^{-i\omega u} - a^{\dag}_-(\omega\hat{x};\beta) e^{i\omega u} \big].
    \label{A4}
\end{equation}
Now, we determine the memory effect operator. As indicated in eq. (\ref{kick}), the memory effect operator is the retarded-time integration of the radiative electric field. Hence,
\begin{equation}
    \int_{-\infty}^{\infty}F_{uz}^{(0)}du = \int_{-\infty}^{\infty}\partial_u A_{z}^{(0)}du = \Delta A_z^{(0)}(z,\bar{z};\beta).
    \label{m_operator}
\end{equation}
From eqs. (\ref{A4}) and (\ref{m_operator}), one can obtain
\begin{equation}
    \Delta A_{z}^{(0)}(z,\bar{z};\beta) = \dfrac{-\sqrt{2}e}{4\pi(1+z\bar{z})} \lim_{\omega\to 0} \omega \big[a_+(\omega\hat{x};\beta) + a^{\dag}_-(\omega\hat{x};\beta) \big].
\end{equation}
This is the electromagnetic memory operator at finite temperatures. Let us now see the action of this memory operator on the ``out-state." Considering a scattering process involving bosons, and assuming that no soft photons are in the out-state, we find,
\begin{equation}
     \braket{\text{out}|\Delta A_z^{(0)}\mathcal{S}|\text{in}} = \dfrac{-\sqrt{2}e}{4\pi(1+z\bar{z})} \lim_{\omega\to 0} \omega \braket{\text{out}|a_{+}(\omega\hat{x};\beta)\mathcal{S}|\text{in}},
    \label{memory_expectation}
\end{equation}
where $\mathcal{S}$ is the scattering matrix.
From eq. (\ref{memory_expectation}), we can see that in the $\omega\to 0$ limit, the thermal memory operator acts on the out-state and creates a soft photon of positive helicity at finite temperatures. Recall the soft photon theorem
\begin{equation}
    \lim_{\omega\to 0}\braket{\text{out}|a_{+}(\omega\hat{x};\beta)\mathcal{S}|\text{in}} = S_{+}^{(0)} \braket{\text{out}|\mathcal{S}|\text{in}} + \mathcal{O}(\omega^0),
\end{equation}
where $S_{+}^{(0)}$ is the leading order soft factor which we have already computed in section (\ref{soft_photon_sec})
\begin{equation}
    S_{+}^{(0)} = \sum_{n} Q_n (p_n\cdot\varepsilon^+) \bigg[ \dfrac{ \eta_n}{p_n\cdot q+i\eta_n \epsilon} - 2\pi i N_B(p_{n0})\delta(p_n\cdot q) \bigg].
    \label{soft_factor}
\end{equation}
Therefore,
\begin{multline}
    \dfrac{\braket{\text{out}|\Delta A_z^{(0)}\mathcal{S}|\text{in}}}{\braket{\text{out}|\mathcal{S}|\text{in}}} = \dfrac{-\sqrt{2}e}{4\pi(1+z\bar{z})} \lim_{\omega\to 0} \omega \sum_{n} Q_n (p_n\cdot\varepsilon^+) \\ \cdot \bigg[ \dfrac{ \eta_n}{p_n\cdot q+i\eta_n \epsilon} - 2\pi i N_B(p_{n0})\delta(p_n\cdot q) \bigg].
    \label{m_boson}
\end{multline}
The second term in the above expression (\ref{m_boson}) contains a Delta function which is non-analytic. This can be regularized using:
\begin{equation}
    \delta(p_n\cdot q) = \dfrac{1}{2\pi i} \bigg(\dfrac{1}{p_n \cdot q - i\epsilon} - \dfrac{1}{p_n \cdot q + i\epsilon} \bigg),
    \label{delta_rep}
\end{equation}
 to get
\begin{multline}
    \dfrac{\braket{\text{out}|\Delta A_z^{(0)}\mathcal{S}|\text{in}}}{\braket{\text{out}|\mathcal{S}|\text{in}}} = \dfrac{-\sqrt{2}e}{4\pi(1+z\bar{z})} \lim_{\omega\to 0} \omega \sum_{n} Q_n \eta_n (p_n\cdot\varepsilon^+) N_B(p_{n0}) \\ \cdot \bigg[ \dfrac{e^{\beta p_{n0}}}{p_n\cdot q+i\eta_n \epsilon} - \dfrac{1}{p_n\cdot q-i\eta_n \epsilon} \bigg].
    \label{memory_boson}
\end{multline}
We note that at zero temperature $(\beta\to\infty)$, the above expression reduces to the expression for the electromagnetic memory effect in zero temperature QED \cite{Pasterski},
\begin{equation}\label{ememory_univ}
    \dfrac{\braket{\text{out}|\Delta A_z^{(0)}\mathcal{S}|\text{in}}}{\braket{\text{out}|\mathcal{S}|\text{in}}} = \dfrac{-\sqrt{2}e}{4\pi(1+z\bar{z})} \lim_{\omega\to 0} \omega \sum_{n} Q_n\eta_n\dfrac{p_n\cdot\varepsilon^+}{p_n\cdot q +i\eta_n\epsilon}.
\end{equation}
 Similar calculations can be done in the case of the scattering of fermions, we obtain the following expression:
\begin{multline}
    \dfrac{\braket{\text{out}|\Delta A_z^{(0)}\mathcal{S}|\text{in}}}{\braket{\text{out}|\mathcal{S}|\text{in}}} = \dfrac{-\sqrt{2}e}{4\pi(1+z\bar{z})} \lim_{\omega\to 0} \omega \sum_{n} Q_n\eta_n(p_n\cdot\varepsilon^+) N_F(p_{n0}) \\ \cdot \bigg[ \dfrac{e^{\beta p_{n0}}}{p_n\cdot q+i\eta_n \epsilon} + \dfrac{1}{p_n\cdot q-i\eta_n \epsilon} \bigg].
    \label{memory_fermion}
\end{multline}
As expected, it depends on the fermionic distribution function and reduces to the universal expression for the memory eq. (\ref{ememory_univ}).\\

\begin{figure}
    \centering
    \includegraphics[scale=0.5]{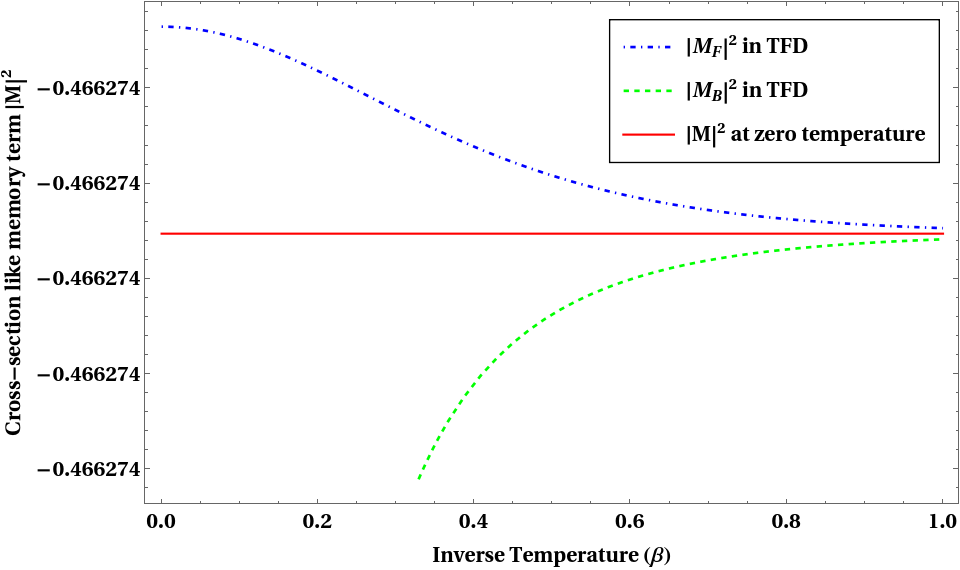}
    \caption{Temperature dependence of the memory expressions ((\ref{F_cross-section}) in dot-dashed line), ((\ref{B_cross-section}) in dashed line), ((\ref{cross-section}) in solid line) at $E=5.0$, $\delta=0.001$, $\theta=0.25\pi$.}
    \label{memory_plot}
\end{figure}

\noindent The memory expressions, (\ref{memory_boson}) and (\ref{memory_fermion}), are formulated in terms of scattering amplitudes. The classical memory expression can be understood as the expectation value of the corresponding memory operator which is a cross-section-like quantity \cite{Strominger_Zhiboedov}. Thus, to analyze the effect of temperature in the memory, we determine the magnitude of the memory expressions, (\ref{memory_boson}) and (\ref{memory_fermion}). For the sake of simplicity, we only consider the contribution from a single external leg in the soft factor. The action of the memory operator on the asymptotic out-state is the residue of the pole in soft factor, thus, we can write
\begin{equation}
    M_F = Q(p\cdot\varepsilon)N_F(p_0) \lim_{\omega\to 0}\omega\bigg(\dfrac{e^{\beta p_0}}{p\cdot q + i\epsilon} + \dfrac{1}{p\cdot q - i\epsilon} \bigg) M_0,
\end{equation}
\begin{equation}
    M_B = Q(p\cdot\varepsilon)N_B(p_0) \lim_{\omega\to 0}\omega\bigg(\dfrac{e^{\beta p_0}}{p\cdot q + i\epsilon} - \dfrac{1}{p\cdot q - i\epsilon} \bigg) M_0.
\end{equation}
Now, we determine the magnitude squared of the above expressions and perform the polarization sum. We use the reference frame in which $p^\mu = (E,0,0,p_3)$, $q^\mu = \omega(1,\sin\theta,0,\cos\theta)$, and hence $p\cdot q = E\omega - p_3\omega\cos\theta$. Further, we consider the high energy limit, i.e. $E \gg m$, $p_3 \approx E$. Therefore,
\begin{equation}
    \sum_{\text{pol}} |M_F|^2 = \dfrac{-Q^2 m^2}{(4E^2 \sin^4{\theta/2}+\delta^2)^2} \bigg(4E^2 \sin^4{\theta/2} + \delta^2 \bigg(\dfrac{e^{\beta E}-1}{e^{\beta E}+1}\bigg)^2 \bigg) |M_0|^2,
    \label{F_cross-section}
\end{equation}
\begin{equation}
    \sum_{\text{pol}} |M_B|^2 = \dfrac{-Q^2 m^2}{(4E^2 \sin^4{\theta/2}+\delta^2)^2} \bigg(4E^2 \sin^4{\theta/2} + \delta^2 \bigg(\dfrac{e^{\beta E}+1}{e^{\beta E}-1}\bigg)^2 \bigg) |M_0|^2.
    \label{B_cross-section}
\end{equation}
Where $\delta = \epsilon/\omega$, and $m$ is the mass of the particle. At zero temperature, i.e. $\beta\to\infty$, we obtain
\begin{equation}
    \sum_{\text{pol}} |M|^2 = \dfrac{-Q^2 m^2}{4E^2 \sin^4{\theta/2}+\delta^2}  |M_0|^2.
    \label{cross-section}
\end{equation}
To understand the effects of temperature-dependent parts on the memory or the soft factors, the expressions in eqs. (\ref{F_cross-section}), (\ref{B_cross-section}), (\ref{cross-section}) are plotted against the inverse of the temperature in fig. (\ref{memory_plot}). The plot shows at finite temperatures, the different scatterers (fermions and bosons)  would make contributions to the memory differently. Scattering involving fermions would be having a decreasing contribution and the same for bosons would be having an increasing contribution as the temperature lowers. Both these curves approach the zero temperature contribution of the memory at very low temperatures. This low-temperature behaviour is also consistent with the expressions given in  (\ref{MTF}) and (\ref{MTB}). It seems, by studying the memory signal at finite temperatures compared to the zero temperature signal one may be able to tell the nature of scatterers involved in the bulk.

\subsection{Gravitational Memory Effect}
\label{GM_sec}
In this section, we study the finite temperature effects on gravitational memory. We follow the same approach as in section (\ref{EM_memory_sec}) and obtain a memory operator for gravitational memory. 

The metric of the asymptotically flat spacetime in the large-$r$ expansion near future null infinity ($\mathcal{I}^+$) in retarded Bondi coordinates is given by \cite{Strominger, Strominger_Zhiboedov, Strominger_2015, Winicour}
\begin{align}
    ds^2 = &-du^2 - 2dudr + 2r^2\gamma_{z\bar{z}}dzd\bar{z} \nonumber \\
    &+\dfrac{2m_B}{r}du^2 + rC_{zz}dz^2 + rC_{\bar{z}\bar{z}}d\bar{z}^2 + D^zC_{zz} dudz + D^{\bar{z}}C_{\bar{z}\bar{z}} dud\bar{z} + .....,
\end{align}
where $m_B(u,z,\bar{z})$ is the Bondi mass aspect, $C_{zz}(u,z,\bar{z})$ describes the time dependence of the gravitational radiation, and $D_z$ is the $\gamma$-covariant derivative. The Bondi news tensor is defined as $N_{zz}=\partial_uC_{zz}$, which determines the energy flux of the radiation. Next, we define memory operator as the retarded time integration of the Bondi news from past of the future null infinity ($\mathcal{I}^{+}_{-}$) to future of the future null infinity ($\mathcal{I}^{+}_{+}$):
\begin{equation}
    \int_{-\infty}^{\infty} N_{zz} du = \int_{-\infty}^{\infty} \partial_u C_{zz}(u,z,\bar{z}) \ du = \Delta C_{zz}(z,\bar{z}).
    \label{G_memory}
\end{equation}
The gravitational field becomes free at late times, $t\to\infty$, and we can approximate it by the mode expansion \cite{Strominger_2015}
\begin{equation}
    h_{\mu\nu}(x)=\kappa \sum_{\alpha=\pm}\int\dfrac{d^3\mathbf{q}}{(2\pi)^3} \dfrac{1}{2\omega} \big[\varepsilon_{\mu\nu}^{\alpha *}(\mathbf{q}) a_\alpha(\mathbf{q}) e^{iq\cdot x} + \varepsilon_{\mu\nu}^{\alpha}(\mathbf{q}) a^{\dag}_\alpha(\mathbf{q}) e^{-iq\cdot x} \big],
\end{equation}
where $\varepsilon^{\pm}_{\mu\nu}$ are the polarization tensors which are traceless, i.e. $\eta^{\mu\nu} \varepsilon^{\pm}_{\mu\nu}=0$, and transverse to the four-momentum of the graviton, i.e. $q^\mu \varepsilon^{\pm}_{\mu\nu}=q^\nu \varepsilon^{\pm}_{\mu\nu}=0$.
The four-momentum of the graviton is parameterized as
\begin{equation}
    q^\mu=\omega\bigg(1, \dfrac{w+\bar{w}}{1+w\bar{w}}, \dfrac{-i(w-\bar{w})}{1+w\bar{w}}, \dfrac{1-w\bar{w}}{1+w\bar{w}} \bigg).
\end{equation}
In this parameterization, the polarization tensors can be written as $\varepsilon^{\pm}_{\mu\nu}=\varepsilon^{\pm}_{\mu} \varepsilon^{\pm}_{\nu}$, where $\varepsilon^{\pm}_{\mu}$ are given by eqs. (\ref{varepsilon_mu+}) and (\ref{varepsilon_mu-}). Let us define the thermal field operator as
\begin{equation}
    h_{\mu\nu}(x;\beta)=\kappa \sum_{\alpha=\pm}\int\dfrac{d^3\mathbf{q}}{(2\pi)^3} \dfrac{1}{2\omega} \big[\varepsilon_{\mu\nu}^{\alpha *}(\mathbf{q}) a_\alpha(\mathbf{q};\beta) e^{iq\cdot x} + \varepsilon_{\mu\nu}^{\alpha}(\mathbf{q}) a^{\dag}_\alpha(\mathbf{q};\beta) e^{-iq\cdot x} \big],
    \label{h_mu_nu1}
\end{equation}
which can be written as
\begin{multline}
    h_{\mu\nu}(x;\beta)=\dfrac{\kappa}{2(2\pi)^3} \sum_{\alpha=\pm}\int_{0}^{\infty} d\omega \ \omega \int d\Omega_{\hat{q}} \big[\varepsilon_{\mu\nu}^{\alpha *}(\mathbf{q}) a_\alpha(\mathbf{q};\beta) e^{-i\omega u - i\omega r(1-\hat{q}\cdot\hat{x})} \\ + \varepsilon_{\mu\nu}^{\alpha}(\mathbf{q}) a^{\dag}_\alpha(\mathbf{q};\beta) e^{i\omega u + i\omega r(1-\hat{q}\cdot\hat{x})} \big].
    \label{h_mu_nu2}
\end{multline}
Now, we take the asymptotic expansion of $h_{\mu\nu}$. As the integrand becomes highly oscillatory in the limit $r\to\infty$, we approximate the above integral by using the stationary phase method employed in (\ref{stationary_phase_identity}). Then eq. (\ref{h_mu_nu2}) reduces to
\begin{equation}
    h_{\mu\nu}(x;\beta) = \dfrac{-i\kappa}{8\pi^2 r} \sum_{\alpha=\pm} \int_{0}^{\infty} d\omega \big[\varepsilon^{\alpha *}_{\mu\nu}(\omega\hat{x}) a_\alpha(\omega\hat{x};\beta) e^{-i\omega u} - \varepsilon^{\alpha}_{\mu\nu}(\omega\hat{x}) a^{\dag}_\alpha(\omega\hat{x};\beta) e^{i\omega u} \big].
    \label{h_asymptotic}
\end{equation}
Next, we perform the following coordinate transformations 
\begin{equation}
    h_{zz}(u,r,z,\bar{z};\beta) = \dfrac{\partial x^\mu}{\partial z} \dfrac{\partial x^\nu}{\partial z} h_{\mu\nu}(x;\beta)
    \label{hzz}
\end{equation}
\begin{equation}
   \therefore \varepsilon^{\alpha *}_{zz} = \dfrac{\partial x^\mu}{\partial z} \dfrac{\partial x^\nu}{\partial z} \varepsilon^{\alpha *}_{\mu\nu}(\omega \hat{x}) = \dfrac{\partial x^\mu}{\partial z} \varepsilon^{\alpha *}_{\mu}(\omega \hat{x}) \ \dfrac{\partial x^\nu}{\partial z} \varepsilon^{\alpha *}_{\nu}(\omega \hat{x}) = \big(\varepsilon^{\alpha *}_{z} \big)^2.
   \label{varepsilon_zz}
\end{equation}
Thus, from eqs. (\ref{varepsilon_zz}), (\ref{varepsilon_z+}) and (\ref{varepsilon_z-}), we obtain
\begin{equation}
    \varepsilon^{+*}_{zz} = \varepsilon^{-}_{zz} = \dfrac{2r^2}{(1+z\bar{z})^2},\,\,
    \varepsilon^{-*}_{zz} = \varepsilon^{+}_{zz} = 0.
\end{equation}
Under the coordinate transformation, eq. (\ref{h_asymptotic}) reduces to
\begin{equation}
    h_{zz}(u,r,z,\bar{z};\beta) = \dfrac{-i\kappa r}{4\pi^2 (1+z\bar{z})^2} \int_{0}^{\infty} d\omega \big[a_{+}(\omega\hat{x};\beta) e^{-i\omega u} - a^{\dag}_{-}(\omega\hat{x};\beta) e^{i\omega u} \big].
\end{equation}
The time dependence of gravitational radiation is described by
\begin{equation}
    C_{zz} = \lim_{r\to\infty} \dfrac{1}{r} h_{zz}.
\end{equation}
Therefore, at $\mathcal{I}^+$,
\begin{equation}
    C_{zz}(u,z,\bar{z};\beta) = \dfrac{-i \kappa}{4\pi^2 (1+z\bar{z})^2} \int_{0}^{\infty} d\omega \big[a_{+}(\omega\hat{x};\beta) e^{-i\omega u} - a^{\dag}_{-}(\omega\hat{x};\beta) e^{i\omega u} \big].
\end{equation}
The Bondi news is defined as $N_{zz}=\partial_u C_{zz}$, therefore
\begin{equation}
    N_{zz}(u,z,\bar{z};\beta) = \dfrac{-\kappa}{4\pi^2 (1+z\bar{z})^2} \int_{0}^{\infty} d\omega \ \omega \big[a_{+}(\omega\hat{x};\beta) e^{-i\omega u} + a^{\dag}_{-}(\omega\hat{x};\beta) e^{i\omega u} \big].
    \label{Nzz}
\end{equation}
From eqs. (\ref{G_memory}) and (\ref{Nzz}), we obtain the gravitational memory operator,
\begin{equation}
    \Delta C_{zz}(z,\bar{z};\beta) = \dfrac{-\kappa}{2\pi(1+z\bar{z})^2} \lim_{\omega\to 0} \omega \big[a_{+}(\omega\hat{x};\beta) + a^{\dag}_{-}(\omega\hat{x};\beta) \big].
\end{equation}
Now, we act the operator $\Delta C_{zz}$ on the asymptotic out-state and obtain
\begin{equation}
    \braket{\text{out}|\Delta C_{zz} \mathcal{S}|\text{in}} = \dfrac{-\kappa}{2\pi(1+z\bar{z})^2} \lim_{\omega\to 0} \omega \braket{\text{out}|a_{+}(\omega\hat{x};\beta) \mathcal{S}|\text{in}}.
    \label{out_DCzz_in}
\end{equation}
The soft graviton theorem reads as
\begin{equation}
    \braket{\text{out}|a_{+}(\omega\hat{x};\beta) \mathcal{S}|\text{in}} = S_{+}^{(0)} \braket{\text{out}|\mathcal{S}|\text{in}} + \mathcal{O}(\omega^0),
    \label{soft_graviton_theorem}
\end{equation}
where $S_{+}^{(0)}$ is the leading order soft factor as determined in section (\ref{soft_graviton_sec}).
Therefore, from eqs. (\ref{out_DCzz_in}), (\ref{soft_graviton_theorem}), and using the regularized form of the delta function (\ref{delta_rep}), we obtain the following expression:
\begin{multline}
    \dfrac{\braket{\text{out}|\Delta C_{zz} \mathcal{S}|\text{in}}}{\braket{\text{out}|\mathcal{S}|\text{in}}} = \dfrac{-\kappa^2}{4\pi(1+z\bar{z})^2} \lim_{\omega\to 0} \omega \sum_{n} \eta_n (p_n\cdot\varepsilon^+)^2 N_B(p_{n0}) \\ \cdot\bigg[\dfrac{e^{\beta p_{n0}}}{p_n\cdot q+i\eta_n\epsilon} - \dfrac{1}{p_n\cdot q-i\eta_n\epsilon} \bigg].
    \label{G_memory_B}
\end{multline}
Similarly, for the scattering of fermions we get, 
\begin{multline}
    \dfrac{\braket{\text{out}|\Delta C_{zz} \mathcal{S}|\text{in}}}{\braket{\text{out}|\mathcal{S}|\text{in}}} = \dfrac{-\kappa^2}{4\pi(1+z\bar{z})^2} \lim_{\omega\to 0} \omega \sum_{n} \eta_n (p_n\cdot\varepsilon^+)^2 N_F(p_{n0}) \\ \cdot\bigg[\dfrac{e^{\beta p_{n0}}}{p_n\cdot q+i\eta_n\epsilon} + \dfrac{1}{p_n\cdot q-i\eta_n\epsilon} \bigg].
    \label{G_memory_F}
\end{multline}
In the zero temperature limit, $\beta\to\infty$, both the expressions (\ref{G_memory_B}) and (\ref{G_memory_F}) reduce to
\begin{equation}
    \dfrac{\braket{\text{out}|\Delta C_{zz} \mathcal{S}|\text{in}}}{\braket{\text{out}|\mathcal{S}|\text{in}}} = \dfrac{-\kappa^2}{4\pi(1+z\bar{z})^2} \lim_{\omega\to 0} \omega \sum_n \dfrac{\eta_n (p_n \cdot \varepsilon^+)^2}{p_n \cdot q + i\eta_n\epsilon},
\end{equation}
which is equivalent to
\begin{equation}
    \dfrac{\braket{\text{out}|\Delta C_{zz} \mathcal{S}|\text{in}}}{\braket{\text{out}|\mathcal{S}|\text{in}}} = \dfrac{-\kappa^2}{4\pi(1+z\bar{z})^2} \varepsilon^+_{\mu\nu} \Bigg(\sum_{\text{out}} \dfrac{p^\mu_j p^\nu_j}{p_j \cdot k} - \sum_{\text{in}} \dfrac{p'^\mu_j p'^\nu_j}{p'_j \cdot k} \Bigg).
\end{equation}
Where, $k^\mu=(1,\hat{q})$. The above expression is related to the Braginsky-Thorne result for the gravitational memory \cite{Strominger_Zhiboedov, BT} which describes the change in the metric at large distances due to the scattering of massive objects such as stars or black holes.

\section{Ward Identity and Soft Theorems}\label{WI}
The ward identity is the statement of the conservation of charges which corresponds to the gauge invariance requirement of the theory. Recently 
 it has been shown to have a direct link with soft theorems. In fact, there can be infinite number of conservation laws established for 'large' gauge transformations at null infinity. The soft theorem emerges as one of the cases of such conservation laws \cite{Strominger}. Here we describe the status of such a picture for soft theorems at finite temperatures. We must note that Ward identities of the form  $q_\mu M^\mu_\text{total}=0$ (see for example (\ref{MT_boson}), lead to the charge conservation. Thermal contributions seem to have no effects on the gauge invariance of soft amplitudes. However, one need to verify this scenario in the set up where asymptotic (gauge) symmetries emerge as an exact symmetry of scattering matrix. 
 
The conserved charges commute with the Hamiltonian of the theory and hence commute with the scattering matrix since
\begin{equation}
    \mathcal{S} \sim \exp{(iHT)}|_{T\to\infty}.
\end{equation}
Therefore, we can write
\begin{equation}
    \braket{\text{out}|[Q_\varepsilon,\mathcal{S}]|\text{in}}=\braket{\text{out}|\big(Q_\varepsilon^+\mathcal{S}-\mathcal{S}Q_\varepsilon^- \big)|\text{in}} = 0. \label{QS}
\end{equation}
Where $Q_\varepsilon$ is the generator for the large gauge transformation with the parameter $\varepsilon$ \cite{Strominger}, i.e.,
\begin{equation}
    \{Q_\varepsilon^+, A_z^{(0)}(u,z,\bar{z}) \} = \partial_z\varepsilon(z,\bar{z}). \label{Q_gen}
\end{equation}
Where
\begin{equation}
    Q_\varepsilon^+ = -2\int d^2z \partial_zN^+ \partial_{\bar{z}}\varepsilon + \int_{\mathcal{I}^+}dud^2z \varepsilon \gamma_{z\bar{z}} j_u^{(2)} \label{Q+},
\end{equation}
and
\begin{equation}
    \partial_zN^+ = \dfrac{-\sqrt{2}}{8\pi e(1+z\bar{z})} \lim_{\omega\to 0} \omega \big[a^{\text{out}}_+(\omega\hat{x}) + a^{\text{out}}_-(\omega\hat{x})^\dag \big]
\end{equation}
is the soft photon mode. Similar expressions can be defined for $\mathcal{I}^-$. Thus, eq. (\ref{QS}) reduces to
\begin{multline}
    -2 \int d^2z \partial_{\bar{z}} \varepsilon \braket{\text{out}|\big(\partial_z N^{+}(z,\bar{z})\mathcal{S}-\mathcal{S}\partial_zN^{-}(z,\bar{z}) \big)|\text{in}} \\= \bigg[\sum_{k=1}^{m} Q_{k}^{\text{in}} \varepsilon(z_{k}^{\text{in}}, \bar{z}_{k}^{\text{in}}) -\sum_{k=1}^{n} Q_{k}^{\text{out}} \varepsilon(z_{k}^{\text{out}}, \bar{z}_{k}^{\text{out}})\bigg] \braket{\text{out}|\mathcal{S}|\text{in}}. \label{Ward_1}
\end{multline}
Where we have assumed $m$ particles come from the asymptotic sphere at $\mathcal{I}^-$ and after the scattering $n$ particles go to the asymptotic sphere at $\mathcal{I}^+$. The above expression indicates an infinite number of Ward identities, one for every $\varepsilon$ \cite{Strominger}. Now, for a particular choice of the large gauge parameter, i.e.,
\begin{equation}
    \varepsilon(w,\bar{w}) = \dfrac{1}{z-w},
\end{equation}
the Ward identity takes the form of the well-known soft photon theorem:
\begin{multline}
    \lim_{\omega\to 0} \omega \braket{\text{out}|\big(a_+^{\text{out}}(\omega \hat{x})\mathcal{S}-\mathcal{S}a^{\text{in}}_-(\omega\hat{x})^\dag \big)|\text{in}} \\= \sqrt{2}e(1+z\bar{z}) \bigg[\sum_{k=1}^{n} \dfrac{Q_{k}^{\text{out}}}{z-z_k^{\text{out}}} -\sum_{k=1}^{m} \dfrac{Q_{k}^{\text{in}}}{z-z_k^{\text{in}}}\bigg] \braket{\text{out}|\mathcal{S}|\text{in}}.
\end{multline}
 Can similar scenario holds for soft amplitudes at finite temperatures? The answer seems to be affirmative.  \\

\noindent To see this, let us introduce the thermal contributions into eq. (\ref{Ward_1})
\begin{multline}
    -2 \int d^2w \partial_{\bar{w}} \varepsilon(w,\bar{w};\beta) \braket{\text{out}|\big(\partial_w N^{+}(w,\bar{w};\beta)\mathcal{S}-\mathcal{S}\partial_wN^{-}(w,\bar{w};\beta) \big)|\text{in}} \\= \bigg[\sum_{k=1}^{m} Q_{k}^{\text{in}} \varepsilon(z_{k}^{\text{in}}, \bar{z}_{k}^{\text{in}};\beta) -\sum_{k=1}^{n} Q_{k}^{\text{out}} \varepsilon(z_{k}^{\text{out}}, \bar{z}_{k}^{\text{out}};\beta)\bigg] \braket{\text{out}|\mathcal{S}|\text{in}}. \label{Ward_T}
\end{multline}
Now, consider a choice of $\varepsilon$ as
\begin{equation}
    \varepsilon(w,\bar{w},\beta) = \dfrac{1}{z-w} + C(\beta) \dfrac{\bar{z}-\bar{w}}{1+w\bar{w}} \delta\bigg(\dfrac{|z-w|^2}{1+w\bar{w}} \bigg),
\end{equation}
where $C(\beta)$ is some function of temperature. Thus, the Ward identity (\ref{Ward_T}) for this choice of $\varepsilon$ takes the following form
\begin{multline}
    4\pi \braket{\text{out}|\big(\partial_z N^{+}(z,\bar{z};\beta)\mathcal{S}-\mathcal{S}\partial_zN^{-}(z,\bar{z};\beta) \big)|\text{in}}
    \\+2C(\beta) \int d^2w \dfrac{\bar{z}-\bar{w}}{1+w\bar{w}} \delta\bigg(\dfrac{|z-w|^2}{1+w\bar{w}} \bigg) \partial_{\bar{w}} \braket{\text{out}|\big(\partial_w N^{+}(w,\bar{w};\beta)\mathcal{S}-\mathcal{S}\partial_wN^{-}(w,\bar{w};\beta) \big)|\text{in}} 
    \\= -\sum_{k=1}^{m+n} \eta_k Q_{k} \bigg[\dfrac{1}{z-z_k} + C(\beta) \dfrac{\bar{z}-\bar{z}_k}{1+z_k\bar{z}_k} \delta\bigg(\dfrac{|z-z_k|^2}{1+z_k\bar{z}_k} \bigg) \bigg] \braket{\text{out}|\mathcal{S}|\text{in}}. \label{Ward_T2}
\end{multline}
The integral on the LHS vanishes which one can simply deduce using the integration by substitution, i.e., by considering $t=\dfrac{|z-w|^2}{1+w\bar{w}}$. Therefore, eq. (\ref{Ward_T2}) reduces to 
\begin{multline}
    \lim_{\omega\to 0} \omega \braket{\text{out}|\big(a_+^{\text{out}}(\omega \hat{x};\beta)\mathcal{S}-\mathcal{S}a^{\text{in}}_-(\omega\hat{x};\beta)^\dag \big)|\text{in}} 
    \\= \sqrt{2}e(1+z\bar{z}) \sum_{k=1}^{m+n} \eta_k Q_{k} \bigg[\dfrac{1}{z-z_k} + C(\beta) \dfrac{\bar{z}-\bar{z}_k}{1+z_k\bar{z}_k} \delta\bigg(\dfrac{|z-z_k|^2}{1+z_k\bar{z}_k} \bigg) \bigg] \braket{\text{out}|\mathcal{S}|\text{in}}.
\end{multline}
The above expression is the finite temperature soft theorem which can be simply verified by comparing the RHS of the above expression in momentum space with the soft factor we obtained in section (\ref{soft_theorem_section}). We note that the temperature dependent function $C(\beta)$ must take the following form: $C(\beta)=\pm 2\pi i N(\beta)$, where $N(\beta)$ is the respective thermal distribution function.

\section{Conclusion}
\label{Conclusion}
\noindent The soft photon and graviton theorems both receive temperature dependent contributions in the leading order soft factors. The factors also depend on the fact whether fermions or bosons are participating in the scattering process, and the universal character of the leading soft factors as known for the zero temperature field theory is getting lost. However, the thermal contributions to the soft factors do not alter the Ward identities. For the soft graviton theorem, this means the equivalence principle is not getting disturbed in a thermal environment when we are in the IR regime. Further, it has been shown, in the low-temperature regime the soft factors will show a somewhat universal character, and the cross-section of the scattering processes would have identical thermal contributions for fermions and bosons. This feature has further been analyzed for memory effects. Using the fact that a time-integrated radiation field at a large distance acts as a memory operator, we showed that memory can be consistently related to the soft factors involving photons and gravitons. The memory effect operator would generate a thermal photon or graviton acting on an asymptotic state.  Next, we plotted the squared amplitudes of soft processes against the inverse temperature to show the difference in contributions of the thermal part for fermionic and bosonic scatterers. At any finite temperature, the soft amplitudes (or cross-sections) involving bosons and fermions contribute opposite to each other and approach the zero temperature amplitude in different ways as it has been shown in fig. (\ref{memory_plot}). This may have an interesting consequence in the context of soft amplitudes. As it turns out the soft processes at finite temperatures would be sensitive to the short-distance details since one in principle can see differences in the cross-section-like quantities for bosonic and fermionic scatterers. We have also shown how the soft theorems at finite temperatures are related to the exact symmetries of scattering matrices through the Ward identities. To establish this we have shown the 'large' gauge parameter needs to be augmented by a thermal part consistent with the soft factor.
\\

Our analysis indicates the thermal effects on soft factors close to zero temperature would be negligible. Therefore, the consequence of a low ambient temperature in the present Universe may not be significant for the soft theorems. However, to draw a conclusion one needs to reformulate the soft theorems for cosmological backgrounds. We have also analyzed the subleading terms for soft theorems at finite temperatures without getting any unexpected results.  It will be interesting to see if the loop corrections to soft amplitudes at finite temperatures generate any unexpected deviation as the loop effects may give rise to some conflicts with the soft limits at finite temperatures.     

\acknowledgments
The authors have benefited from useful discussions with
Sudipta Sarkar and Dileep Jatkar. D.S. thanks Subhodeep Sarkar for help in preparing the draft. S.B. thanks Avirup Ghosh for the discussions. The authors also acknowledge the infrastructural support from their host institute IIIT Allahabad. S.B. is partially supported by SERB-DST through MATRICS grant MTR/2022/000170.

\end{document}